\documentclass[showpacs]{revtex4}
\usepackage[dvips]{graphicx}

\begin{document}

\title{Creation of macroscopic superpositions of flow states with Bose-Einstein condensates}

\author{Jacob Dunningham}

\affiliation{School of Physics and Astronomy, University of Leeds, Leeds LS2 9JT, UK}
\author{David Hallwood}
\affiliation{Clarendon Laboratory, Department of Physics, University of Oxford, Oxford OX1 3PU, UK}

\begin{abstract}
We present a straightforward scheme for creating macroscopic superpositions of different superfluid flow states of Bose-Einstein condensates trapped in optical 
lattices. This scheme has the great advantage that all the techniques required are achievable with current experiments. Furthermore, the relative difficulty of creating cats scales favorably with the size of the cat. This means that this scheme may be well-suited to creating superpositions involving large numbers of particles. Such states may have interesting technological applications such as making quantum-limited measurements of angular momentum.

\end{abstract}

\pacs{03.75.Lm, 03.75.Gg, 03.75.Kk}

\maketitle

\section{Introduction}

Quantum mechanics allows objects to exist in a coherent superposition of different states. Th does not depend on the size of the system and so offers the fascinating prospect of being able to create superpositions of macroscopically distinct states.
Multiparticle superposition states have been observed in a number of systems including three photons \cite{Mitchell2004a}, C${}_{60}$ molecules \cite{Arndt1999a}, and the internal state of four ${}^{9}$Be${}^{+}$ ions \cite{Sackett2000a}.
Experimental signatures of larger scale quantum phenomena were shown when Rouse \emph{et al.}~\cite{rouse_95} observed resonant tunneling between two macroscopically distinct states in a superconducting quantum interference device (SQUID). The observed superposition was between states of 
different flux or opposite currents flowing around a loop. These currents consisted of approximately $10^9$ Cooper pairs, meaning tunneling between two macroscopically distinct states had been achieved. Similar systems have also been used to create macroscopic superpositions of these two supercurrent states \cite{friedman_00, wal_00}. 

Bose-Einstein condensates (BECs) are a promising system for realizing similar results. They are composed of up to $10^{9}$ atoms with a high proportion in the same quantum state and are
sufficiently cold to undergo a quantum phase transition from a superfluid to a 
Mott insulator~\cite{greiner_02}. 
There have already been a number of theoretical proposals for producing cat states with BECs in a range of different set-ups ~\cite{cirac,Gordon1999a,Ruostekoski1998a,Dunningham2001a,Hallwood2006a}. 

In this paper, we present a scheme for producing a macroscopic superposition of different superfluid flow states (or equivalently angular quasi-momentum states) in a ring of coupled BECs \cite{Hallwood2006a}.
This is important because it provides a direct manifestation of quantum mechanics at the macroscopic level in a new system. As discussed by Leggett \cite{leggett_02}, such states are important for testing the limits of validity of quantum mechanics. All the steps in our scheme are straightforward and should be achievable with present experiments. The BEC system may also have significant advantages over SQUIDs since it is highly controllable: the coupling between condensates and the strength of the interactions between atoms can be tuned over many orders of magnitude. This scheme has the added advantage that the degree of control over the system that is required to create a cat state scales favorably with the number atoms involved. This suggests that such a scheme may be well-suited to creating `large' cats.
Finally, we discuss how a macroscopic superposition of different superfluid flows may also be of technological interest. One possibility is quantum-limited measurements of angular momentum or, equivalently, ultra-precise gyroscopes \cite{dubetsky2006a}.

\section{The scheme}

\begin{figure}[b]
\includegraphics[width=7.0cm]{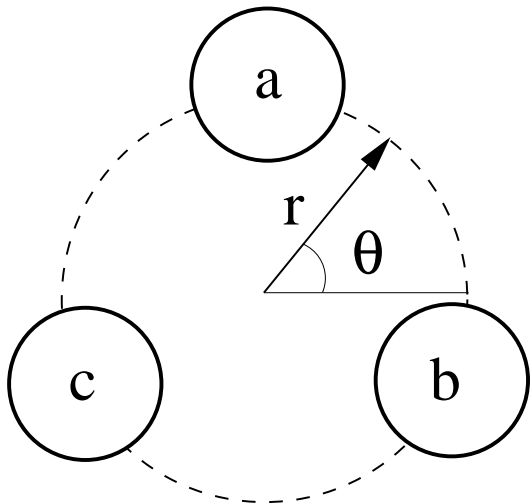}
\caption{Diagram of the set-up. Three BECs respectively denoted $a$, $b$ and $c$ are trapped in an optical lattice in a ring geometry. It is convenient to describe this set-up in polar coordinates $(r,\theta)$.} \label{threesite}
\end{figure}

The system we consider consists of condensed atoms trapped by the optical dipole force in a lattice formed by a standing wave of laser light. In particular, we shall consider three lattice sites in a ring geometry (see Figure 1). Each of the lattice sites is coupled to its neighbors by quantum mechanical tunneling through the potential barriers separating them.

This system can be described by the Bose-Hubbard Hamiltonian \cite{Jaksch}
\begin{equation}
H = -J\left(a^{\dag}b + b^{\dag}c + c^{\dag}a + \rm{h.c.} \right) + \frac{U}{2}\left( a^{\dag}{}^{2}a^{2} + b^{\dag}{}^{2}b^{2} +
 c^{\dag}{}^{2}c^{2} \right), \label{bosehubb}
\end{equation}
where $a$, $b$, and $c$ are the annihilation operators for an atom at the respective lattice sites.
The strength of the tunneling between sites, $J$, can be adjusted in
experiments by changing the intensity of the standing wave, thus altering the potential barrier between sites. The interaction
strength between atoms, $U$, is at best only weakly dependent on the potential, but can be controlled by using Feshbach resonances.
In the scheme presented here, we will be concerned only with changing the value of $J$.

Using Bloch's theorem, which describes the wave function of a particle in a periodic potential, we can write the wave function of a Bose-Einstein condensate 
in polar coordinates as,
\begin{equation}
\psi(r,\theta+2\pi n/3) = e^{i2\pi \ell n/3}\psi(r,\theta), \label{Bloch}
\end{equation}
where $n$ is an integer and $\ell$ is some number to be determined.  We can interpret $\ell$ as being the angular quasi-momentum of the mode in units of $\hbar$. Normally when using Bloch's theorem (e.g. to describe electrons moving in a periodic lattice) periodic boundary conditions are put in by hand as an approximation to simplify the analysis. In the geometry considered here, periodic boundary conditions are true physical constraints needed to ensure that the wave function 
is single-valued everywhere. This means that the description of this system using Bloch states is exact.  The boundary condition can be written as,
\begin{equation}
\psi(r,\theta+2\pi) = \psi(r,\theta),
\end{equation}
and we see from (\ref{Bloch}) that this requires that $\ell$ is an integer. Another way of say this is that the angular momentum must be quantized in units of $\hbar$.
Moreover, we see from (\ref{Bloch}) that changing $\ell$ by an integer multiple of 3 does not change the wave function. This means that a complete basis is formed by restricting $\ell$ to the values $\ell\in \{0,\pm 1\}$. 
The annihilation operators for this complete quasi-momentum basis are,
\begin{eqnarray}
\alpha &=& \frac{1}{\sqrt{3}}\left( a+b+c \right) \label{alphamode} \\
\beta &=& \frac{1}{\sqrt{3}}\left( a+ e^{i2\pi/3}b+ e^{i4\pi/3}c \right) \label{betamode} \\
\gamma &=& \frac{1}{\sqrt{3}}\left( a+ e^{-i2\pi/3}b+ e^{-i4\pi/3}c \right). \label{gammamode}
\end{eqnarray}

Let us now consider the details of the cat creation process.
The great advantage of the scheme presented here is that it is straightforward to carry out and within reach of present experiments. Macroscopic superpositions can be created simply by starting with the ground state in the lattice when the barriers are low, rapidly increasing the intensity of the optical lattice, and then waiting for a certain time before rapidly decreasing the intensity of the optical lattice again.

For simplicity we begin by considering a system of only three atoms \cite{footnote} (we will consider larger numbers later in the paper). We 
want to create a state of the form,
\begin{equation}
|\psi\rangle = \frac{1}{\sqrt{3}}\left( |3,0,0\rangle_{\alpha\beta\gamma} + |0,3,0\rangle_{\alpha\beta\gamma} +
|0,0,3\rangle_{\alpha\beta\gamma} \right), \label{threecat}
\end{equation}
i.e. a superposition of all the atoms having no angular momentum and all having one unit of angular momentum clockwise and all having one unit of angular momentum anticlockwise.

\begin{figure}[t]
\includegraphics[width=10.0cm]{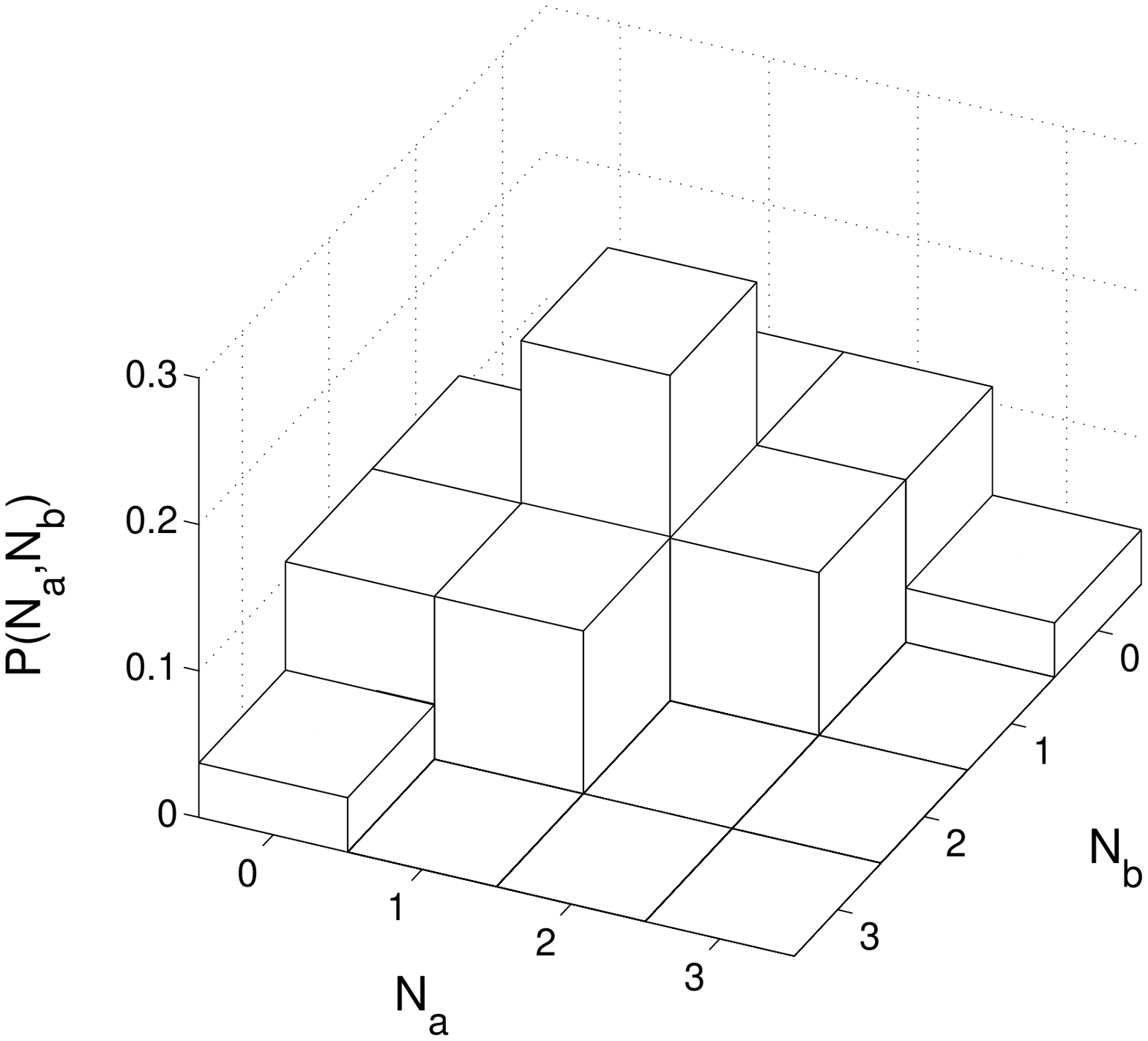}
\caption{Probability distribution of the number of atoms in lattice sites $a$ and $b$ for the initial ground state of the system, i.e. all atoms in the $\alpha$ quasi-momentum mode, for $N=3$ (\ref{threeground}).} \label{flow1}
\end{figure}

\begin{figure}[b]
\includegraphics[width=10.0cm]{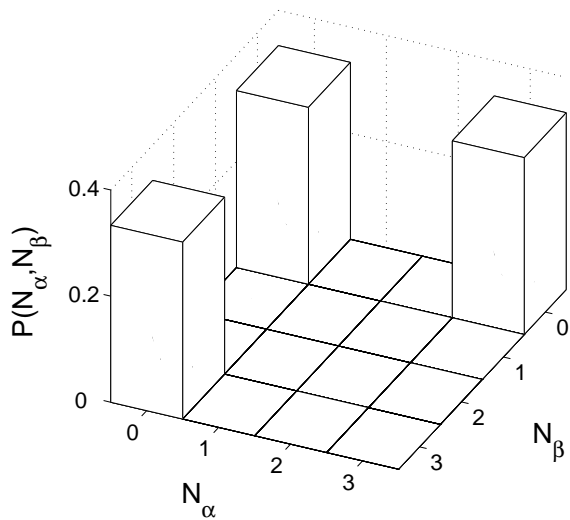}
\caption{Final number distribution of atoms in the $\alpha$ and $\beta$ modes after the cat creation process when $N=3$.} \label{flow2}
\end{figure}

The Hamiltonian for the condensate in the lattice when the potential barriers are sufficiently low that the energy associated with tunneling dominates the 
energy associated with interactions, $J\gg U$, is given approximately by 
\begin{equation}
H = -J\left(a^{\dag}b + b^{\dag}c + c^{\dag}a + \rm{h.c.} \right) = -J\left(2\alpha^{\dag}\alpha - \beta^{\dag}\beta - \gamma^{\dag}\gamma\right).   \label{hamlow}
\end{equation}
The eigenstates of this Hamiltonian are the quasi-momentum modes (\ref{alphamode}) - (\ref{gammamode}) and the ground state is 
when all the atoms are in mode $\alpha$ (i.e. the zero quasi-momentum Bloch state). For three atoms, this can be written in the basis of the number of atoms per lattice site as,
\begin{eqnarray}
|\psi\rangle &=& \frac{1}{9\sqrt{2}}\left(a^{\dag}+b^{\dag} +c^{\dag}\right)^{3}|0,0,0\rangle \nonumber\\
&=& \frac{1}{3\sqrt{3}}\left( |3,0,0\rangle +|0,3,0\rangle + |0,0,3\rangle \right) + \frac{\sqrt{2}}{3}|1,1,1\rangle \nonumber \\
&+& \frac{1}{3}\left( |1,2,0\rangle +|2,1,0\rangle + |1,0,2\rangle +|2,0,1\rangle +|0,1,2\rangle +|0,2,1\rangle \right). \label{threeground}
\end{eqnarray} 
Figure~2 shows the number distribution of atoms in the lattice sites,
$P(N_{a},N_{b}) = \left|\langle N_{a},N_{b},3-N_{a}-N_{b}|\psi\rangle\right|^{2}$, for this ground state. As we might expect, we see that it is relatively unlikely to find all the atoms in a single lattice site, but more likely to find them equally distributed between sites.

Our claim is that, if we start with a state of the form of (\ref{threeground}), we can create a cat state of the form of (\ref{threecat}) by 
first rapidly increasing the intensity of the optical lattice, then waiting for a certain time, and finally rapidly decreasing the light intensity again.
We now study the details of this process.

The first step is to rapidly increase the intensity of the trapping laser light, thereby increasing the height of the potential barriers between the sites and so decreasing the coupling, $J$. We want the rate at which this change is made to be much greater than rate of tunneling between lattice sites, but still adiabatic with respect to the spacing of energy levels at each site.
This separation of timescales is possible and has been experimentally demonstrated for condensates in an optical lattice \cite{Greiner2002b}. If the coupling between the sites is decreased sufficiently by the increase in intensity that the coupling energy is small compared with the energy due to the interactions between atoms, the Hamiltonian can be written approximately as,
\begin{equation}
H= \frac{U}{2}\left( a^{\dag}{}^{2}a^{2} + b^{\dag}{}^{2}b^{2} + c^{\dag}{}^{2}c^{2} \right). \label{hamhigh}
\end{equation}
Immediately after the barriers have been raised, the state is given by (\ref{threeground}) since it has not had time to evolve. 
If we now allow this state to evolve under the influence of the Hamiltonian (\ref{hamhigh}), after time $t$, the state is,
\begin{eqnarray}
|\psi\rangle_{2} &=& \frac{1}{3\sqrt{3}}e^{i3Ut}\left( |3,0,0\rangle +|0,3,0\rangle + |0,0,3\rangle \right) + \frac{\sqrt{2}}{3}|1,1,1\rangle \nonumber \\
&+& \frac{1}{3}e^{iUt}\left( |1,2,0\rangle +|2,1,0\rangle + |1,0,2\rangle +|2,0,1\rangle +|0,1,2\rangle +|0,2,1\rangle \right). \label{Ustate}
\end{eqnarray}

We now want to see whether this evolves to a cat state. For a state of the form of (\ref{threecat}), we need the the probability for all three atoms to be found in
mode $\alpha$, $P_{\alpha}(3)$, to be one-third and similarly for the probabilities for modes $\beta$, $P_{\beta}(3)$ and $\gamma$, $P_{\gamma}(3)$. Calculating these probabilities directly we get,
\begin{eqnarray}
P_{\alpha}(3) &=& \frac{1}{3^4}\left[ 41 +24\cos(Ut) +12\cos(2Ut) + 4\cos(3Ut)\right]   \label{Palpha} \\
P_{\beta}(3) &=& P_{\gamma}(3) = \frac{1}{3^4}\left[ 41 -12\cos(Ut) -6\cos(2Ut) + 4\cos(3Ut)\right].  \label{Pbeta}
\end{eqnarray}
If we pick the evolution time to be $t=2\pi/(3U)$, we get $P_{\alpha}(3)=P_{\beta}(3)=P_{\gamma}(3) = 1/3$.
This is precisely what we want: an equal superposition of all the atoms in $\alpha$, and all in $\beta$, and all in $\gamma$.
A plot of the full number distribution in the quasi-momentum basis is shown in Figure~3 and confirms that a cat state of the form of (\ref{threecat}) has been created.

Finally we rapidly lower the potential barriers again. As before, this is done much faster than the rate of tunneling between the potential wells, but slowly with respect to the spacing between energy levels within each well. This leaves us with the state (\ref{threecat}) \cite{footnote2} evolving due to the Hamiltonian (\ref{hamlow}). Since this Hamiltonian has no terms that couple different quasi-momentum modes, the population in each does not change with time and so the cat-like structure of the state is preserved.  
The relative phases between terms, however, do change with time since the different quasi-momentum modes have different energies, i.e. the non-rotating mode, $\alpha$, has a lower energy than the rotating modes $\beta$ and $\gamma$.

\section{Larger numbers}

We have now demonstrated how the cat creation process works for three atoms. In order for this to be a truly powerful technique, we need to demonstrate that it also works for larger numbers of particles, $N>3$, to create states of the form,
\begin{equation}
|\psi\rangle = \frac{1}{\sqrt{3}}\left( |N,0,0\rangle_{\alpha\beta\gamma} + |0,N,0\rangle_{\alpha\beta\gamma} +
|0,0,N\rangle_{\alpha\beta\gamma} \right). \label{Ncat}
\end{equation}
For simplicity, let us suppose that the total number of atoms in the system is a multiple of three, i.e. $N=3n$, where $n$ is an integer. The initial ground state of the condensed atoms trapped in the lattice when the energy of the coupling term dominates, $J \gg U$, is
\begin{eqnarray}
|\psi\rangle &=& \frac{1}{\sqrt{3^{N}N!}}(a^{\dag}+b^{\dag}+c^{\dag})^{N}|0,0,0\rangle \nonumber \\
&=& \frac{1}{\sqrt{3^{N}}}\sum_{p=0}^{N}\sum_{q=0}^{N-p}\sqrt{\frac{N!}{p!q!(N-p-q)!}}|p,q,N-p-q\rangle.
\end{eqnarray}
The probability distribution of the number of atoms in lattice sites $a$ and $b$ is plotted in Figure~4 for $N=30$. 

\begin{figure}[t]
\includegraphics[width=10.0cm]{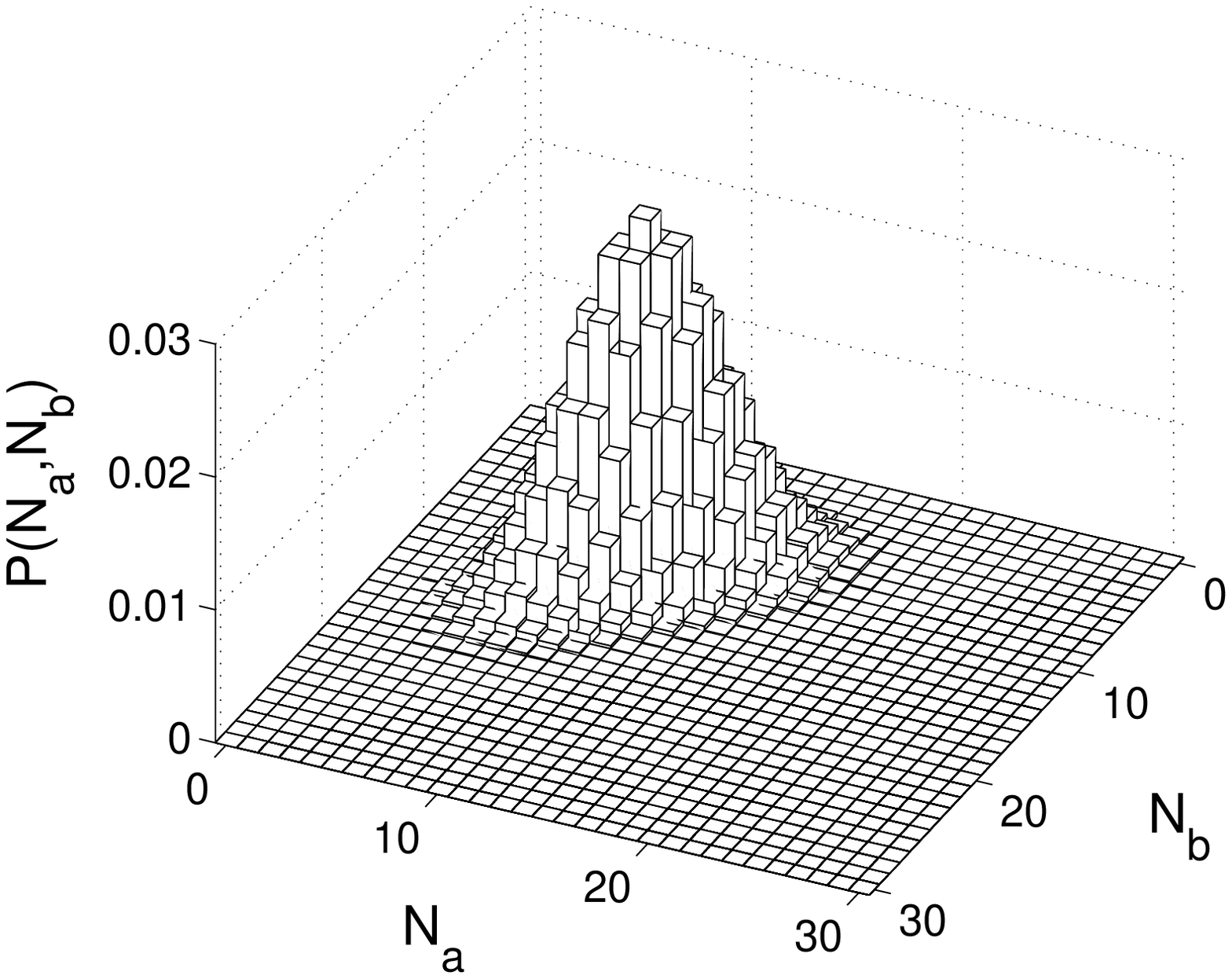}
\caption{Probability distribution of the number of atoms in lattice sites $a$ and $b$ for the initial ground state of the system, i.e. all atoms in the $\alpha$ quasi-momentum mode, for $N=30$.} \label{bigflow1}
\end{figure}

\begin{figure}[b]
\includegraphics[width=10.0cm]{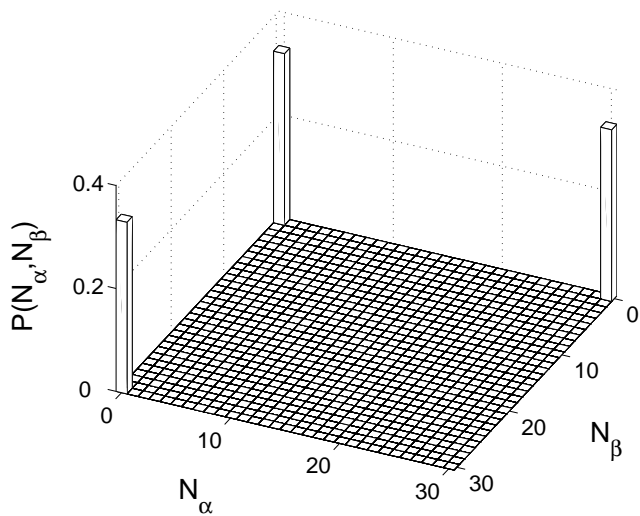}
\caption{Final number distribution of atoms in the $\alpha$ and $\beta$ modes after the cat creation process when $N=30$.} \label{bigflow2}
\end{figure}

We now follow exactly the same procedure as for three atoms: rapidly raise the potential barriers, hold for time $t=2\pi/(3U)$, and then rapidly reduce the barriers to their initial height. It is straightforward to calculate the probabilities for all the atoms to be in a single quasi-momentum mode and, as before, we get $P_{\alpha}(N) = P_{\beta}(N) = P_{\gamma}(N)=1/3$. The full number distribution is shown in Figure~5 for $N=30$ and confirms that a cat state of the form (\ref{Ncat}) has been created.
This is a great result - it means that we do not need to know the number of atoms in the initial state. So long as the total number of atoms is a multiple of three, we get (\ref{Ncat}).

We would now like to consider what happens when $N\neq 3n$. To achieve this, it is helpful to define a measure of `cattiness', $C$, i.e  a measure of how similar our final state is to (\ref{Ncat}). 
For this, we choose \cite{footnote3},
\begin{equation}
C= 3\left[ P_{\alpha}(N)P_{\beta}(N)P_{\gamma}(N)\right]^{1/3}. \label{catmeasure}
\end{equation}
This can take values ranging from zero, when the state is very different from (\ref{Ncat}), up to one, when the state has the same form as (\ref{Ncat}). 
In Figure~6, we have plotted the value of $C$ for the final state (created by the procedure outlined above) for a range of values of $N$. We see that we get a state of the form of (\ref{Ncat}) if $N$ is a multiple of three, but not otherwise \cite{footnote4}.
This, however, should not overly concern us. 
For one, this is a very simple procedure for creating cat states. The straightforward implementation of this scheme far outweighs the fact that it only works for one in three trials, which is certainly not a prohibitively low success rate for usefulness in a range of schemes. Secondly, it is easy to post-select trials where the cat was successfully created and throw away all the unsuccessful attempts.

\begin{figure}[h]
\includegraphics[width=10.0cm]{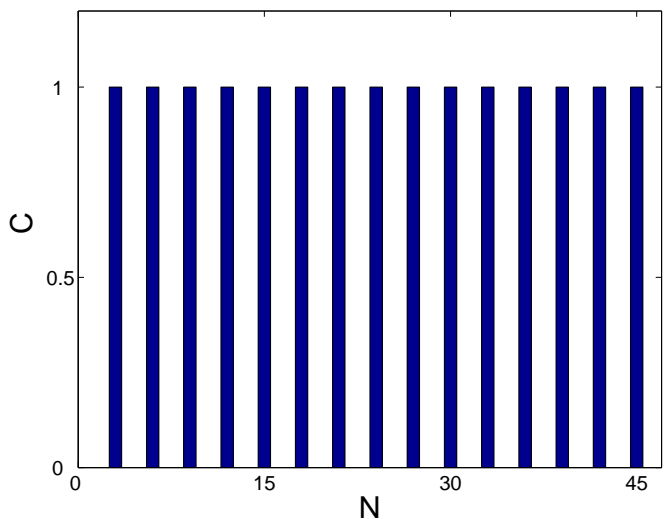}
\caption{Plot of the `cattiness' measure $C$ defined by (\ref{catmeasure}) as a function of the total number of atoms in the system, $N$. A cat of the form of (\ref{Ncat}) is created (i.e. $C=1$) only if $N$ is a multiple of three.} \label{figure6}
\end{figure}

\section{Discussion}

One of the great advantages of this scheme over a procedure we previously proposed \cite{Hallwood2006a} is that, instead of having to control the rate of rotation of the lattice to a high degree of precision, we need only control the time that the system is allowed to evolve when the potential barriers are high. This should be much easier to achieve in practice. In this section, we investigate how sensitive the cat creation process is to the accuracy of this timing.

As our benchmark we will take a value of $C=0.9$. This corresponds (in the case of the state being symmetric in $\alpha$, $\beta$, and $\gamma$) to $90\%$ of the atoms being in a state of the form of (\ref{Ncat}). We would now like to investigate how accurately we need to control the time of the nonlinear evolution in order to achieve a value of $C\geq 0.9$ for the final state. 
If we take the time of the nonlinear evolution to be $t=(1+\delta)2\pi/(3U)$, where $\delta$ is the error in the timing, we would  like to find the maximum value of $\delta$ such that $C\geq 0.9$ as a function of $N$. We denote this maximum value as $\delta_{0}$.

A plot of these results is shown in Figure~7. The crosses show the points for total numbers of atoms that are a multiple of three and the solid curve is a best-fit line. The approximate scaling from this fit is,
\begin{equation}
\delta_{0} \approx 0.24/N. \label{scaling}
\end{equation}
This means that the error that can be tolerated in the timing scales as $\Delta t \sim 1/UN$. We note that this result is consistent with the calculation in the previous section for three particles. In (\ref{Palpha}) and (\ref{Pbeta}), the most rapidly varying terms are proportional to $\cos(3Ut)$, which means that the error in $t$ scales as $\Delta t\sim 1/(3U)$.

Here we see the great advantage of this scheme: the accuracy with which we need to control the timing scales as $1/N$. This is much more favorable than the exponential dependence on $N$ that the rate of rotation of the lattice had to be controlled in a previous scheme \cite{Hallwood2006a}. 
This means that the present scheme has a relatively low premium on creating cat states with large numbers and so is particularly well-suited to this purpose. This is good news since the advantages of cat states in quantum schemes and investigating fundamental issues of physics are more fully exploited the larger the cat state.

The fact that the non-linear evolution needs to be controlled more accurately the larger the cat is will ultimately limit the size of cats that can be created by this process. However, this fact could also be turned to an advantage. For example, we could use this scheme to measure the value of $U$ accurately - something that is difficult to achieve experimentally. One way this could be achieved is to carry out the procedure outlined above for different values of $t$ and then make measurements on the final state. 
Whenever, $Ut$ is a multiple of $2\pi/3$, a cat of the from of (\ref{Ncat}) will be created when $N$ is a multiple of three. This means that one-third of the time we would find all the atoms in the same angular momentum mode - a highly unlikely event if a cat state of the form of (\ref{Ncat}) had not been created. This, then, provides a clear means of calibrating $Ut$ to within $1/N$ and, since time can be measured extremely accurately, would enable a precise determination of $U$.

\begin{figure}[h]
\includegraphics[width=10.0cm]{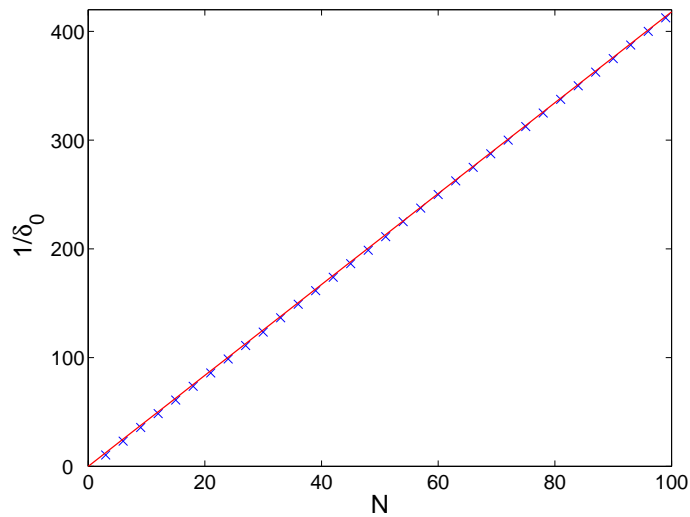}
\caption{Plot of $1/\delta_{0}$ as a function of $N$, where $\delta_{0}$ is the maximum value of the error in the timing of the nonlinear evolution, $\delta$, that allows $C\geq 0.9$. The crosses are numerically calculated data points for values of $N$ that are multiples of three. The solid curve is a line of best fit intended to find the scaling. The solid curve should not be interpreted as an interpolation between points, since the cat creation process only works when $N$ is a multiple of three.} \label{figure7}
\end{figure}

\section{Read-out and interferometry} \label{interferometer}

One interesting application of macroscopic superpositions of flow states is that they may allow for quantum-limited measurements of angular momentum, i.e ultraprecise gyroscopes \cite{Hallwood2006a,dubetsky2006a}. To achieve this, we would like to carry out a form of interferometry using cat states. A similar idea makes use of `quantum beam splitters' to accurately measure phase shifts in optical and double-well BEC systems \cite{Kim2005a, Dunningham2006a}. 

The basic idea is to create a macroscopic superposition of superfluid flows using the procedure outlined above. The cat state is then held for some time $\Delta t$ and any rotation of the lattice during this time will shift the energy levels of the different angular quasi-momenta and so encode phases on the state. The cat-making procedure is then repeated, in analogy with the second beam splitter in a Mach-Zehnder interferometer, and the number of atoms in each angular momentum mode is measured. This should enable the rotation of the lattice to be determined accurately.

In the present discussion, we will limit ourselves to the case where the total number of atoms is a multiple of three. In this case, it can be shown that the transformation of the state by the cat creation process is,
\begin{equation}
\left( \begin{array}{c} |N,0,0\rangle_{\alpha\beta\gamma} \\
|0,N,0\rangle_{\alpha\beta\gamma} \\
|0,0,N\rangle_{\alpha\beta\gamma} \end{array} \right)
\longrightarrow
U \left( \begin{array}{c} |N,0,0\rangle_{\alpha\beta\gamma} \\
|0,N,0\rangle_{\alpha\beta\gamma} \\
|0,0,N\rangle_{\alpha\beta\gamma} \end{array} \right),
\end{equation}
where $U$ is the unitary transformation,
\begin{equation}
U=\frac{1}{\sqrt{3}}\left( \begin{array}{ccc} e^{-i2\pi/3} & 1 & 1 \\
1 & e^{-i2\pi/3} & 1  \\
1 & 1 & e^{-i2\pi/3}  \end{array} \right).
\end{equation}
It is straightforward to show that $U^{3} = {\bf 1}$. This means that, since that cat creation process corresponds to the operator $U$, we pick the second cat operation (corresponding to the second `beam splitter') to be $U^{2}$. This ensures that the second operation is the inverse of the first.
Experimentally this could be achieved simply by allowing the state to evolve twice as long with the non-linear interactions, i.e. $t=4\pi/(3U)$ \cite{footnote5}.

We now need to consider how rotations of the lattice shift the energy of $\alpha$, $\beta$, and $\gamma$. For small changes in the angular momentum, $\Delta L$, we can write,
\begin{equation}
\Delta E \approx \frac{\partial E}{\partial L} \Delta L = \xi L,
\end{equation}
where $\xi$ is given by $L/I$ and $I$ is the moment of inertia of the atoms. We saw earlier that the angular momenta of $\alpha$, $\beta$, and $\gamma$ are $0$, $\hbar$, and $-\hbar$ respectively. This means that the respective energy shifts of these modes are $0$, $\xi\hbar$, and $-\xi\hbar$ and the Hamiltonian describing the system at this stage can be modified from (\ref{hamlow}) to give,
\begin{equation}
H=\hbar\left[ -2J\alpha^{\dag}\alpha +(J+\xi)\beta^{\dag}\beta +(J-\xi)\gamma^{\dag}\gamma \right].
\end{equation}
The matrix describing the phase shift of each mode due to evolution with this Hamiltonian for time $\Delta t$, $e^{iH\Delta t/\hbar}$, can be written in the basis
$\{|N,0,0\rangle_{\alpha\beta\gamma}, |0,N,0\rangle_{\alpha\beta\gamma}, |0,0,N\rangle_{\alpha\beta\gamma}\}$ as,
\begin{equation}
Q=\left( \begin{array}{ccc} e^{-i3NJ\Delta t} & 0 & 0 \\
0 & e^{iN\xi\Delta t} & 0  \\
0 & 0 & e^{-iN\xi\Delta t}  \end{array} \right),
\end{equation}
where an overall phase has been neglected.

Before discussing the results, let us first summarize the interferometry scheme. We first create a cat by the process described earlier and then hold it for some time, $\Delta t$. It is during this time that any slight changes in the rotation of the lattice are detected. Finally, we repeat the cat creation process, but this time for double the duration of the nonlinear interaction, and detect which angular momentum modes the atoms of the final state are in. 
Even though we have two cat-creation processes in this scheme, the overall chance of success in any given trial is still one-third. This is because either both processes are successful or both are not depending on whether the total number of particles in the system is a multiple of three, i.e. the success of the two processes are not independent. Furthermore, we can throw away any unsuccessful trials since we will know that it has been unsuccessful when we measure it if all the atoms are not found in the same angular momentum mode, $\alpha$, $\beta$, or $\gamma$. This means that the restriction that the total number of atoms must be a multiple of three should not overly concern us.

The final state created by this scheme, if we begin with the ground state of the system (i.e. all the atoms in the non-rotating $\alpha$ mode), is $|\psi\rangle_{\rm final} = U^{2}QU|N,0,0\rangle_{\alpha\beta\gamma}$. When $\xi \Delta t=0$, i.e. there is no rotation of the lattice or the time over which the rotation is applied is vanishingly small, we get $Q={\bf 1}$ and the final state is the same as the initial state, as it should be. The probabilities for finding all the atoms in $\alpha$, $\beta$, and $\gamma$ respectively are,
\begin{eqnarray}
P_{\alpha} &=& \frac{1}{9}\left[ 1 +4\cos^{2}(N\xi\Delta t) + 4\cos(N\xi\Delta t)\cos(3NJ\Delta t)\right]  \label{Pfalf}\\
P_{\beta} &=& \frac{1}{9}\left[ 1 +4\cos^{2}(N\xi\Delta t-2\pi/3) + 4\cos(N\xi\Delta t -2\pi/3)\cos(3NJ\Delta t)\right] \label{Pfbet}\\
P_{\gamma} &=& \frac{1}{9}\left[ 1 +4\cos^{2}(N\xi\Delta t + 2\pi/3) + 4\cos(N\xi\Delta t + 2\pi/3)\cos(3NJ\Delta t)\right], \label{Pfgam}
\end{eqnarray}
and $P_{\alpha} + P_{\beta} + P_{\gamma} = 1$, as it should when $N$ is a multiple of three.

We see that the frequency of oscillation of the probabilities (\ref{Pfalf}) -- (\ref{Pfgam}) scales as $N$. So by changing $\xi\Delta t = L\Delta t/I$ by an amount of the order of $1/N$ all the atoms can be  
transferred from one angular momentum mode to another. This is a dramatic observable that should be able to be measured experimentally, thereby enabling  measurements of angular momentum with a precision that scales as $1/N$. This is the same scaling as the fundamental quantum limit (Heisenberg limit) and, since $N$ can be large for macroscopic superpositions, opens up the exciting possibility of achieving ultra-precise gyroscopes.

\section{Conclusion}

We have presented a detailed scheme for creating macroscopic superpositions of different superfluid flow states in an optical lattice. Such states are of fundamental interest in exploring the boundary between the regimes where quantum and classical physics are valid. The scheme we have proposed has some pleasing features. BECs in optical lattices present a system free from impurities that is highly controllable -- the coupling between sites and the interactions between atoms can be adjusted over several orders of magnitude. Furthermore, this scheme is very simple and well within reach of present experiments. Macroscopic superposition states can be created simply by rapidly switching the intensity of the laser beams that form the optical lattice. This scheme has the great advantage that there is a relatively low premium on creating superpositions with large numbers of atoms. The timing of the nonlinear interaction needs to be controlled to an accuracy that scales as $1/N$, which is much more favorable than the degree of control required in other schemes. This suggests that this scheme may present a viable route to achieving macroscopic superpositions of superfluid flow states in the laboratory. Along with enabling detailed studies of quantum mechanics, these states may have interesting technological applications including ultra-precise quantum-limited gyroscopes.

\section{Acknowledgements}
This work was supported by the UK Engineering and Physical Sciences Research Council (Grant No. GR/S99297/01).


\begin{thebibliography}{99}
\bibitem{footnote} We shall use the term `condensate' throughout this paper to refer to atoms in the ground state of the trapping potential even though this may sometimes refer to only a small number of atoms.
\bibitem{footnote2} There may be additional phase factors between the terms in the superposition, but these are not of particular interest in the present discussion.
\bibitem{footnote3} The (perhaps obvious) choice of using fidelity as this measure is not appropriate here as we are not interested in the phases between terms in the superpostion.
\bibitem{footnote4} Interestingly, macroscopic superposition states are still created when $N\neq 3n$, however they are not of the form of (\ref{Ncat}).
\bibitem{footnote5} An alternative would be to reverse the sign of the non-linear interactions for the second operation, for example by making use of Feshbach resonances. Of course this would be more difficult to achieve in practice than simply allowing the system to evolve for twice as long.
\end{thebibliography}
\end{document}